\documentstyle[12pt,epsfig]{article}

\catcode`\@=11
\long\def\@makefntext#1{
\protect\noindent \hbox to 3.2pt {\hskip-.9pt
$^{{\ninerm\@thefnmark}}$\hfil}#1\hfill}                %CAN BE USED

\def\@makefnmark{\hbox to 0pt{$^{\@thefnmark}$\hss}}  %ORIGINAL

\def\ps@myheadings{\let\@mkboth\@gobbletwo
\def\@oddhead{\hbox{}
\rightmark\hfil\ninerm\thepage}
\def\@oddfoot{}\def\@evenhead{\ninerm\thepage\hfil
\leftmark\hbox{}}\def\@evenfoot{}
\def\sectionmark##1{}\def\subsectionmark##1{}}

\renewcommand{\thefootnote}{\fnsymbol{footnote}}

\def\sectionc{\@startsection {section}{1}{\z@}{-3.5ex plus -1ex minus 
    -.2ex}{2.3ex plus .2ex}{\bf }}
\def\subsectionc{\@startsection{subsection}{2}{\z@}{-3.25ex plus -1ex minus 
   -.2ex}{1.5ex plus .2ex}{\it }}
\renewcommand{\section}[1]{\sectionc{#1}\hspace*{\parindent}}
\renewcommand{\subsection}[1]{\subsectionc{#1}\hspace*{\parindent}}
\newcounter{appendixc}
\newcounter{subappendixc}[appendixc]
\newcounter{subsubappendixc}[subappendixc]

\renewcommand{\appendix}[1] {\vspace*{0.6cm}
        \refstepcounter{appendixc}
        \setcounter{figure}{0}
        \setcounter{table}{0}
        \setcounter{equation}{0}
        \renewcommand{\thefigure}{\Alph{appendixc}.\arabic{figure}}
        \renewcommand{\thetable}{\Alph{appendixc}.\arabic{table}}
        \renewcommand{\theappendixc}{\Alph{appendixc}}
        \renewcommand{\theequation}{\Alph{appendixc}.\arabic{equation}}
        \noindent{\bf Appendix \theappendixc #1}\par\vspace*{0.4cm}}

\def\abstracts#1{{
        \centering{\begin{minipage}{13.2truecm}\footnotesize\baselineskip=13pt\noindent
        \parindent=0pt #1
        \end{minipage}}\par}}

\renewenvironment{thebibliography}[1]
        {\begin{list}{\arabic{enumi}.}
        {\usecounter{enumi}\setlength{\parsep}{0pt}
\setlength{\leftmargin 0.75cm}{\rightmargin 0pt}
         \setlength{\itemsep}{0pt} \settowidth
        {\labelwidth}{#1.}\sloppy}}{\end{list}}

\newcounter{itemlistc}
\newcounter{romanlistc}
\newcounter{alphlistc}
\newcounter{arabiclistc}

\newcommand{\fcaption}[1]{
        \refstepcounter{figure}
        \setbox\@tempboxa = \hbox{\footnotesize Figure~\thefigure. #1}
        \ifdim \wd\@tempboxa > 6in
           {\begin{center}
        \parbox{6in}{\footnotesize\baselineskip=13pt Figure~\thefigure. #1}
            \end{center}}
        \else
             {\begin{center}
             {\footnotesize Figure~\thefigure. #1}
              \end{center}}
        \fi}

\newcommand{\tcaption}[1]{
        \refstepcounter{table}
        \setbox\@tempboxa = \hbox{\footnotesize Table~\thetable. #1}
        \ifdim \wd\@tempboxa > 6in
           {\begin{center}
        \parbox{6in}{\footnotesize\baselineskip=13pt Table~\thetable. #1}
            \end{center}}
        \else
             {\begin{center}
             {\footnotesize Table~\thetable. #1}
              \end{center}}
        \fi}

\def\@citex[#1]#2{\if@filesw\immediate\write\@auxout
        {\string\citation{#2}}\fi
\def\@citea{}\@cite{\@for\@citeb:=#2\do
        {\@citea\def\@citea{,}\@ifundefined
        {b@\@citeb}{{\bf ?}\@warning
        {Citation `\@citeb' on page \thepage \space undefined}}
        {\csname b@\@citeb\endcsname}}}{#1}}

\newif\if@cghi
\def\cite{\@cghitrue\@ifnextchar [{\@tempswatrue
        \@citex}{\@tempswafalse\@citex[]}}
\def\citelow{\@cghifalse\@ifnextchar [{\@tempswatrue
        \@citex}{\@tempswafalse\@citex[]}}
\def\@cite#1#2{{$\null^{#1}$\if@tempswa\typeout
        {IJCGA warning: optional citation argument
        ignored: `#2'} \fi}}

 1
 1
 1

\font\ninerm=cmr9

\begin{document}

\begin{flushright}
ADP-96-31/T230
\end{flushright}
\centerline{\normalsize\bf Variation of hadron masses in finite nuclei}
\baselineskip=15pt

%\vfill
\vspace*{0.6cm}
\centerline{\footnotesize Koichi SAITO}
\baselineskip=13pt
\centerline{\footnotesize\it Physics Division, Tohoku College of Pharmacy}
\baselineskip=13pt
\centerline{\footnotesize\it Sendai 981, Japan}
\centerline{\footnotesize E-mail: ksaito@nucl.phys.tohoku.ac.jp}
\vspace*{0.3cm}
%\centerline{\footnotesize and}
\vspace*{0.3cm}
\centerline{\footnotesize Kazuo TSUSHIMA and Anthony W. THOMAS}
\baselineskip=13pt
\centerline{\footnotesize\it Department of Physics and Mathematical Physics, 
University of Adelaide}
\baselineskip=13pt
\centerline{\footnotesize\it Adelaide, SA 5005, Australia}

%\vfill
\vspace*{0.6cm}
\abstracts{
Using a self-consistent, Hartree description for both infinite nuclear 
matter and finite nuclei based on a relativistic quark model (the 
quark-meson coupling model), we investigate the variation of 
the masses of the non-strange vector mesons, the hyperons and the nucleon 
in infinite nuclear matter and in finite nuclei.  
}

%\vspace*{0.6cm}
\normalsize\baselineskip=15pt
\setcounter{footnote}{0}
\renewcommand{\thefootnote}{\alph{footnote}}

\section{Introduction}\label{sec:intro}  
One of the most interesting future directions in nuclear physics may 
be to 
study how nuclear matter properties change as the environment 
changes. Forthcoming ultra-relativistic heavy-ion experiments are 
expected to give significant information on the strong interaction 
(i.e., QCD ) in matter, through the detection of changes in hadronic 
properties. 

In particular, the variations in hadron masses in finite nuclei 
have attracted wide interest because such changes 
could be a signal of the formation of hot hadronic and/or quark-gluon 
matter in energetic nucleus-nucleus collisions\cite{exp/ko}.  
Several authors have studied the vector-meson ($\omega$, 
$\rho$, $\phi$) masses using the vector dominance model, 
QCD sum rules and the Walecka model (QHD), and have reported that 
these masses 
decrease in the nuclear medium\cite{hat1}. 

Using a self-consistent, Hartree description for both nuclear 
matter and finite nuclei in the 
quark-meson coupling (QMC) model, we report on the variation of 
the masses of the non-strange vector mesons, the hyperons and the nucleon 
in finite nuclei.  

\section{The QMC model}\label{sec:qmc}
The QMC model may be viewed as an extension of QHD in which the 
nucleons still interact through the exchange of $\sigma$ and $\omega$ 
mesons.  However, the mesons couple not to point-like nucleons but to
confined u and d quarks.  In studies of infinite nuclear 
matter\cite{qmc} it was found that
the extra degrees of freedom provided by the internal structure of the
nucleon mean that one gets quite an acceptable value for the
incompressibility once the coupling constants are chosen to
reproduce the correct saturation energy and density for symmetric 
nuclear matter. This is a significant improvement on QHD 
at the same level of sophistication. 
There have been several interesting applications to the properties of 
finite nuclei using the local-density approximation\cite{appl}.  

In Ref.\cite{fnt}, we extended the QMC model to 
finite nuclei in the Born-Oppenheimer approximation.  
The QMC model not only reproduces the saturation properties of nuclear 
matter but also describes fairly well 
the observed charge density distributions and neutron density distributions 
of static, closed-shell nuclei from $^{16}$O to $^{208}$Pb (see also 
Ref.\cite{tsushim}).  
The basic result in the QMC model is that, 
in the scalar ($\sigma$) and vector ($\omega$) 
meson fields, the nucleon behaves essentially as a point-like 
particle with an effective mass 
$M_N^{\star}$ (which depends on the position only through the $\sigma$ 
field) moving in a vector potential generated by the $\omega$ 
meson. 

To calculate the variation of hadron masses in finite nuclei, it 
is necessary to include the effect of not only the changes in
nucleon structure but 
also those of the mesons in the model.  
Because of the vector character, the vector interactions have {\em no effect 
on the hadron structure} except for an overall phase in the wave function, 
which gives a shift in the hadron energy.  Therefore, our effective 
Lagrangian density in mean field approximation takes the form:
\begin{eqnarray}
{\cal L}_{QMC}&=& \overline{\psi} [i \gamma \cdot \partial 
- M_N  + g_\sigma (\sigma({\vec r})) \sigma({\vec r}) 
- g_\omega \omega({\vec r}) \gamma_0 \nonumber \\
&-& g_\rho \frac{\tau^N_3}{2} b({\vec r}) \gamma_0 
- \frac{e}{2} (1+\tau^N_3) A({\vec r}) \gamma_0 ] \psi \nonumber \\
&-& \frac{1}{2}[ (\nabla \sigma({\vec r}))^2 + 
m_{\sigma}^{\star}(\sigma)^2 \sigma({\vec r})^2 ] 
+ \frac{1}{2}[ (\nabla \omega({\vec r}))^2 + m_{\omega}^{\star}(\sigma)^2 
\omega({\vec r})^2 ] \nonumber \\
&+& \frac{1}{2}[ (\nabla b({\vec r}))^2 + m_{\rho}^{\star}(\sigma)^2 
b({\vec r})^2 ] 
+ \frac{1}{2} (\nabla A({\vec r}))^2, 
\label{qmclag}
\end{eqnarray}
where we notice that the vector meson masses depend on only 
the $\sigma$ field at the point ${\vec r}$ in the nucleus.  

\section{A new scaling phenomenon in hadron masses}\label{sec:scaling}
In the QMC model the changes in the hadron masses can be 
described in terms of the common scalar ($\sigma$) field in nuclei (see also 
Ref.\cite{hadron}).  
Supposing that a strange quark does not couple to the scalar field directly, 
the decrease of the hadron mass is well approximated by 
\begin{equation}
\delta M_j^{\star} \equiv M_j - M_j^{\star} \simeq \frac{n_0}{3} (g_{\sigma} 
\sigma(r)) \left[ 1 - \frac{a_j}{2} (g_{\sigma} \sigma(r)) \right], 
\label{mas1}
\end{equation}
where $M_j(M_j^{\star})$ is the hadron mass ($j$ = N, $\omega, \rho, \Lambda$, 
etc) in free space (matter), $n_0$ is the number of 
non-strange quarks in the hadron and $a_j$ is a constant 
depending on the hadron structure -- it ranges around $8.5 \sim 9.5 \times 
10^{-4}$ (MeV$^{-1}$) for N, $\omega, \rho, \Lambda$ and $\Xi$.  
At low density we find that the mass reduction is
\begin{equation}
\frac{M_j^{\star}}{M_j} \simeq 1 - \frac{n_0}{3} \left[ \frac{g_{\sigma} 
\sigma(r)}{M_j} \right], 
\ \ \ \mbox{and} \ \ \ 
g_{\sigma} \sigma \simeq 200 (\mbox{MeV}) \left( \frac{\rho_B}{\rho_0}
\right), 
\label{mas2}
\end{equation}
and we then get 
\begin{eqnarray}
\frac{M_N^{\star}}{M_N} &\simeq& 1 - 0.2 \left( \frac{\rho_B}{\rho_0} \right), 
\ \ \ 
\frac{m_{\omega}^{\star}}{m_{\omega}} \simeq 1 - 0.16 \left( \frac{\rho_B}
{\rho_0} \right), 
\nonumber \\
\frac{M_{\Lambda}^{\star}}{M_{\Lambda}} \approx 
\frac{M_{\Sigma}^{\star}}{M_{\Sigma}} &\simeq& 1 - 0.11 \left( 
\frac{\rho_B}{\rho_0} \right), 
\ \ \ 
\frac{M_{\Xi}^{\star}}{M_{\Xi}} \simeq 1 - 0.05 \left( 
\frac{\rho_B}{\rho_0} \right). 
\label{mas3}
\end{eqnarray}
Since the 
dependence of the constant $a_j$ on the hadrons is weak (in Eq.(\ref{mas1})), 
our results lead to  
a new scaling relation describing the variation of hadron masses with
density: 
\begin{equation}
\frac{\delta M_{\omega,\rho}^{\star}}{\delta M_N^{\star}} \approx 
\frac{\delta M_{\Lambda}^{\star}}{\delta M_N^{\star}} \approx \frac{2}{3}, 
\ \ \ 
\frac{\delta M_{\Xi}^{\star}}{\delta M_N^{\star}} \approx \frac{1}{3}, 
\ \mbox{ etc}.
\label{mas4}
\end{equation}
This scaling is relevant over the range of $\rho_B$ up to $\sim 3\rho_0$. 

\section{Hadron masses in finite nuclei}\label{sec:hadron}
\begin{figure}[htb]
\begin{center}
%\rule{7cm}{0.2mm}
%\vskip 1.0cm
%\rule{7cm}{0.2mm}
\epsfig{file=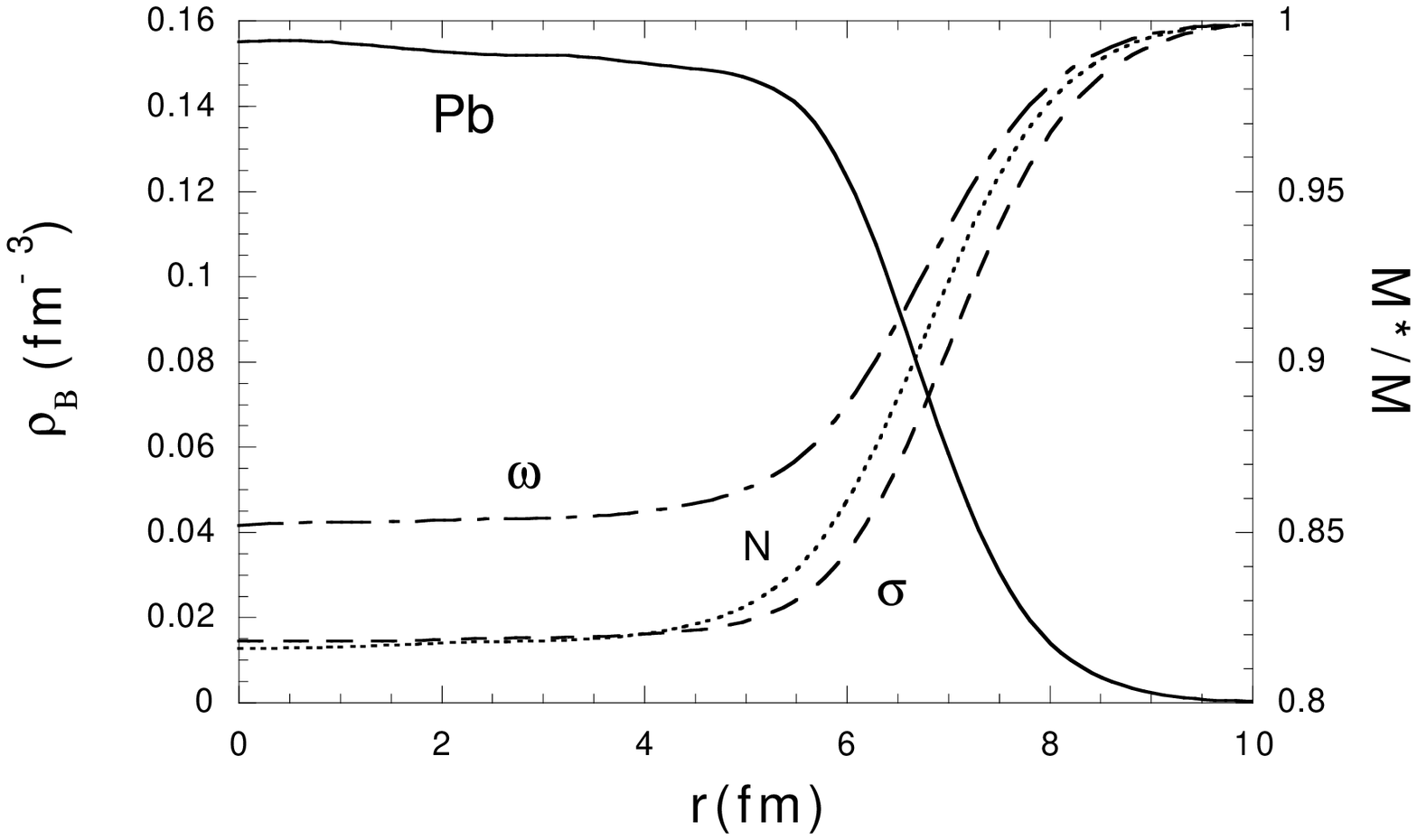,height=7cm}
\end{center}
\fcaption{The variation of nucleon, $\omega$ and $\sigma$ meson masses 
in $^{208}$Pb.  The nuclear baryon density, $\rho_B$, is also shown.  The 
right (left) scale is for the effective mass (the baryon density). 
\label{fig:varmas}}
\end{figure}

Using the effective Lagrangian density, Eq.(\ref{qmclag}), we have performed 
some self-consistent calculations of the variation of hadron masses in 
finite nuclei -- 
for example, the initial result for $^{208}$Pb is shown in 
Fig.\ref{fig:varmas} 
(we will present detailed numerical studies elsewhere).  

In this model it is, however, hard to deal with the change 
of the $\sigma$-meson mass in medium because it couples strongly 
to the pseudoscalar ($\pi$) channel, which requires a direct 
treatment of chiral symmetry in medium\cite{hat2}.  
Here we suppose that $m_\sigma^{\star}$ in Eq.(\ref{qmclag}) is also 
given by the same form as 
those of the vector mesons -- see Eq.(\ref{mas1}).  

As shown in the figure the $\omega$-meson mass is reduced 
about 15\% while the nucleon and the $\sigma$-meson masses decrease about 
20\% in the interior of Pb, which are also expected in other 
various models\cite{exp/ko,hat1}.  

\section{Summary}\label{sec:sum}
We have studied the spectral change of the hadrons in both nuclear matter 
and finite nuclei using the extended QMC model. 
As several authors have 
suggested\cite{hat1}, the hadron mass is 
reduced due to the change of the scalar mean-field in medium.  
In the 
present model the hadron mass can be related to the number of 
non-strange quarks and the strength of the scalar mean-field.  The 
hadron masses are simply connected to one another, and the 
relationship among them is given by Eq.(\ref{mas4}), 
which is effective over a wide range of the nuclear density.  
Furthermore, we have shown the variation of the hadron masses in lead. 
It will be quite important to perform such calculations for finite nuclei 
in order to interpret the results observed in 
forthcoming ultra-relativistic heavy-ion experiments.  

This work was supported by the Australian Research Council.


\vspace*{0.6cm}

\begin{thebibliography}{99}
%
\bibitem{exp/ko} HELIOS-3 collaboration, {\it Nucl. Phys.} {\bf A590} 
(1995) 93c. \\
CERES collaboration, {\it ibid.} {\bf A590} (1995) 103c. \\
G.Q. Li, C.M. Ko and G.E. Brown, nucl-th/9608040 (1996).
%
\bibitem{hat1} For summary, T. Hatsuda, nucl-th/9608037 (1996).
%
\bibitem{qmc} P.A.M. Guichon, {\it Phys. Lett.} {\bf B200} (1988) 235. \\
K. Saito and A.W. Thomas, {\it Phys. Lett.} {\bf B327} (1994) 9. 
%
\bibitem{appl} K. Saito and A.W. Thomas, {\it Phys. Lett.} {\bf B335} 
(1994) 17. \\
K. Saito and A.W. Thomas, {\it Nucl. Phys.} {\bf A574} (1994) 659. \\
K. Saito and A.W. Thomas, {\it Phys. Lett.} {\bf B363} (1995) 157. 
%
\bibitem{fnt} P.A.M. Guichon {\it et al}., 
{\it Nucl. Phys.} {\bf A601} (1996) 349. \\
K. Saito {\it et al}., Univ. of Adelaide preprint 
ADP-96-17/T220 (nucl-th/9606020, 1996), to appear in {\it Nucl. Phys.} 
{\bf A}. 
%
\bibitem{tsushim} K. Tsushima {\it et al}., these Proceedings.
%
\bibitem{hadron} K. Saito and A.W. Thomas, {\it Phys. Rev.} 
{\bf C51} (1995) 2757.
%
\bibitem{hat2} T. Hatsuda and T. Kunihiro, {\it Phys. Rep.} {\bf 247} (1994) 
221.
%
\end{thebibliography}
\end{document}